\documentclass[11pt,a4paper,american]{article}
\usepackage{jheppub} 

\usepackage{amssymb}
\usepackage{amsmath}
\usepackage{xcolor}
\usepackage{graphicx}
 \usepackage{floatrow}
 \usepackage{soul}

\makeatletter

\pdfpageheight\paperheight
\pdfpagewidth\paperwidth

\numberwithin{equation}{section}

\pdfoutput=1
\renewcommand\[{\begin{equation}}
\renewcommand\]{\end{equation}}

\makeatother

\usepackage{babel}

\begin{document}

\title{Emergence of ghosts in Horndeski theory}

\author{Eugeny~Babichev} 
\affiliation{Universit\'e Paris-Saclay, CNRS/IN2P3, IJCLab, 91405 Orsay, France}

\abstract{
We show that starting from initial conditions with stable perturbations, evolution of a galileon scalar field results in the appearance of a ghost later on. 
To demonstrate this, we consider a theory with k-essence and cubic galileon Lagrangians on a fixed Minkowski background.
Explicit analytical solutions of simple waves are constructed, on top of which  
a healthy scalar degree of freedom is shown to be converted onto a ghost.
Such a transformation is smooth and moreover perturbations remain hyperbolic all the time (until a caustic forms). 
We discuss a relation between the ghost and the appearance of closed causal curves for such solutions.
}

\maketitle


\section{Introduction}

In various modifications of gravity it is not unusual that the stability of solutions depends on both a theory and a particular solution at hand. 
In particular, when studying small perturbations on top of a background solution, one finds that for the same theory perturbations 
may be stable for one solution and be unstable for another. 
Horndeski theory~\cite{Horndeski} (see also~\cite{Deffayet:2011gz,Kobayashi:2011nu}) is an example of such a behaviour. 
By construction, the equations of motion of the theory are of the second order. This ensures the absence of extra ghost degrees of freedom, which would result in the Ostrogradsky instability~\cite{Ostrogradski}. 
In spite of the absence of an extra pathological degree of freedom, the existing scalar and tensor degrees of freedom are not always healthy. 
For a given theory, one background solution may contain a ghost (or gradient) instability and another background for the same theory may be stable. 

One of the motivations for this work is to understand whether a solution with no pathologies can dynamically evolve into a solution with a ghost in the spectrum of perturbations. In other words, the question we ask is whether it is possible that a healthy degree of freedom is converted into a ghost during a smooth dynamical process. 
For this purpose we consider a scalar field theory living in Minkowski space time with non-dynamical metric. 
The action of the theory is a sum of k-essence~\cite{ArmendarizPicon:1999rj} and a simple cubic galileon~\cite{Nicolis:2008in}. 
To study non-trivial dynamics of the theory, we use the construction of propagating waves in Horndeski theory~\cite{Babichev:2016hys} (see also \cite{Tanahashi:2017kgn,Pasmatsiou:2017vcw,deRham:2016ged}).
It was shown in~\cite{Babichev:2016hys} that for general initial conditions, solutions for dynamical simple wave in 1+1 result in caustic formation for non-standard kinetic terms (with the exception for the DBI-like Lagrangian~\cite{Mukohyama:2016ipl}). 
An interesting feature of simple wave solutions in k-essence is that they are not affected by the presence of higher-order galileon terms, i.e. a simple wave solution of pure k-essence theory is also a solution of a theory with the same k-essence part of the Lagrangian plus any higher order (cubic, quartic and quintic) galileons. 

We use this particular feature to study perturbations of a theory with k-essence and cubic galileon terms in the action, on top of a simple wave.
We show that before reaching a caustic singularity in the solution, initially healthy perturbations of the galileon may become ghost-like. 
Whether the scalar degree of freedom evolves into ghost, depends on the background solution. 
However, for any choice of the free parameter of the cubic Lagrangian, one can organise such initial conditions 
that an originally healthy solution turns into an unstable solution plagued with a ghost. 

The plan of the paper is the following. In the next section we write down the action and the equations of motion and review simple wave solutions. 
In the Section~\ref{sec:perturbations} we study perturbations on top of these solutions. In Section~\ref{sec:ghosts} we show, using the results of the previous section, how a ghost emerges in the spectrum of perturbations, once the solution approaches a caustic singularity. 
We conclude and discuss in Section~\ref{sec:conslusions}.

\section{The background solution}
\label{sec:background}
We consider the scalar field dynamics derived from the following Lagrangian,
\begin{equation}
\label{L}
\mathcal{L} =  \mathcal{K}(X) + \frac{\gamma}2 X \Box\phi,
\end{equation}
where the first term in the action is the k-essence term, with $\mathcal{K}$ being a non-linear function of the standard kinetic term $X\equiv \frac12(\partial_\mu\phi\partial^\mu\phi)$; 
and the second term is the cubic galileon with the constant coefficient $\gamma$ that may be positive or negative.

We will assume that the metric is non-dynamical and flat, 
so the only dynamical variable in the theory is the scalar field.
In addition we restrict ourselves to the case of two-dimensional motion, i.e. 
$\phi$ is a function of the time coordinate $t$ and one spatial coordinate $x$. 
As it has been shown in~\cite{Babichev:2016hys}, in this case one can construct exact non-trivial solutions of propagating waves, called simple waves~\cite{Courant}.
Because of the choice of the solution and the symmetry of the problem (i.e. the assumption that the solutions depend only on one spacial coordinate and time)
the generalised cubic, quartic and quintic galileon terms  do not affect the dynamics of the scalar field. 
I.e. simple wave solutions are solely determined by the k-essence term $\mathcal{K}(X)$.
For quartic and quintic galileons this statement is true for any $\phi(t,x)$, even off-shell. 
This straightforwardly follows from the form of the Lagrangian in terms of the fully antisymmetric tensor and 
the fact that we consider dynamics in $1+1$ dimensions.
On the other hand the cubic galileon or its generalised version does not affect the dynamics of the scalar field {\it on a simple wave solution}, while for more general solutions this is no longer true. 
A detailed discussion of this peculiar behaviour can be found in~\cite{Babichev:2016hys}.
Thus, setting $\gamma=0$, the equation of motion for the scalar field obtained by the variation of~(\ref{L}) with respect to $\phi$ reads,
\begin{equation}
\label{eom0}
	\left( \mathcal{K}_X g^{\mu\nu}+\mathcal{K}_{XX}\nabla^\mu\phi\nabla^\nu\phi \right)\nabla_\mu\nabla_\nu\phi 	=0,
\end{equation}
where the subscript denotes the derivative with respect to $X$, i.e. $\mathcal{K}_X\equiv d\mathcal{K}/dX$,  $\mathcal{K}_{XX}\equiv d^2\mathcal{K}/(dX)^2$.

It is known that although k-essence is a non-linear theory, it admits travelling wave solutions, which propagate with the speed of light keeping the same shape~\cite{Babichev:2007dw}. 
This also holds for full  Horndeski theory with dynamical metric~\cite{Babichev:2012qs}.
However, more general initial conditions lead to more complicates wave-like solutions.
For generic initial conditions satisfying a simple wave solution, caustics are formed during the evolution of the scalar field~\cite{Babichev:2016hys}\footnote{Note that one can construct an extension of the k-essence theory to include a second dynamical field, so that the formation of caustics can be avoided~\cite{Babichev:2017lrx,Babichev:2018twg}. For such a theory, the low-energy modes effectively behave as k-essence, while at high energy the kinetic structure becomes canonical and thus caustic-free.}. 
In this paper we will use the construction of exact solutions for simple waves presented in~\cite{Babichev:2016hys}.
We will show that depending on the theory and the background solution, perturbations of the scalar field may become ghost-like before caustics are formed,
even though for the initial configuration perturbations are healthy. 

Let us briefly recall the approach and results of~\cite{Babichev:2016hys} on simple wave solutions. 
Because the action (\ref{L}) is shift-symmetric, i.e. it is invariant under the change $\phi\to\phi+$const., 
it is convenient to work with the following variables $\tau$ and $\chi$, 
\begin{equation}\label{defp}
	\tau = \dot \phi, \ \ \ \chi = \phi',
\end{equation}
In particular, using these new variables the kinetic term is expressed as $X=\frac12(\tau^2-\chi^2)$.
By the use of~(\ref{defp}) the equation of motion~(\ref{eom0}) can be written in the  form,
\begin{equation}\label{eom1}
	\left(\mathcal{K}_X +\tau^2 \mathcal{K}_{XX}\right) \dot\tau  -2\, \tau \chi \mathcal{K}_{XX} \tau' + \left(-  \mathcal{K}_X + \chi^2 \mathcal{K}_{XX}\right) \chi' =0.
\end{equation}
One can study the above equation~(\ref{eom1}) by the method of characteristics~\cite{Courant,Vladimirov}.
Provided that the equation is hyperbolic, i.e. 
\begin{equation*}
	\label{hyper}
	\mathcal{K}_X^2 + 2X \mathcal{K}_X \mathcal{K}_{XX} >0\, ,
\end{equation*}
there are two families of characteristics,
\begin{equation}\label{xipm}
	\xi_{\pm}= \frac{-\tau\chi \mathcal{K}_{XX}\pm  \sqrt{\mathcal{K}_X^2+2X\mathcal{K}_X\mathcal{K}_{XX}}}{\mathcal{K}_X+\tau^2 \mathcal{K}_{XX}},
\end{equation}
where each $\xi$ corresponds to the slope of the characteristic curve, $\xi \equiv (dx/dt)_\sigma = x_\sigma/t_\sigma$, with a parameter $\sigma$ along the curve 
(for details see~\cite{Babichev:2016hys}).
It is then convenient to introduce the following quantity, 
\begin{equation}\label{cs}
c_s^2 = \left(1+2X\frac{\mathcal{K}_{XX}}{\mathcal{K}_X}\right)^{-1},
\end{equation}
which coincides with the propagation speed of small perturbations on a background solution with timelike $\partial_\mu\phi$~\cite{Garriga:1999vw}. 
Using the above definition (\ref{cs}), the expression for characteristics~(\ref{xipm}) can be simplified as follows,
\begin{equation}\label{xipm2}
	\xi_\pm= \frac{\pm \tau c_s -\chi}{\tau \mp \chi c_s},
\end{equation}
which has a clear physical interpretation as a relativistic velocity-addition formula~\cite{Babichev:2016hys}.

An advantage of this approach involving the method of characteristics is that 
the partial differential equation~(\ref{eom1}) can be written as 
a system of four ordinary differential equations on dependent variables $t$, $x$, $\tau$ and $\chi$ 
as functions of two independent variables $\sigma_+$ and $\sigma_-$,
\begin{eqnarray}
	&& \frac{dx}{d\sigma_\pm} = \xi_\pm \frac{dt}{d\sigma_\pm},   \label{char1}\\
	 && \xi_\pm \left(\mathcal{K}_X +\tau^2 \mathcal{K}_{XX}\right)  \frac{d\tau}{d\sigma_\pm} + \left(-  \mathcal{K}_X + \chi^2 \mathcal{K}_{XX}\right) \frac{d\chi}{d\sigma_\pm} =0, \label{char2} 
\end{eqnarray}
where $\sigma_\pm$ are the parameters along the characteristics $\xi_\pm$. 
Note that Eq.~(\ref{char1})  can be viewed as the definition of characteristics.
In our case the equations~(\ref{char2}) decouple from the other two equations~(\ref{char1}), 
since $t$ and $x$ do not enter~(\ref{char2}) explicitly. 

From~(\ref{eom1}) and (\ref{char2}) one can derive a relation that is crucial for construction of simple wave solutions, 
\begin{equation}\label{charG}
	\Gamma_\pm = -\xi_\mp,
\end{equation}
where $\Gamma_{\pm} \equiv \left(d\tau / d\chi \right)_{\pm} $ are the characteristics in the $(\tau,\chi)$ plane. 

\begin{figure}[t]
\includegraphics[angle=-90,width=\textwidth]{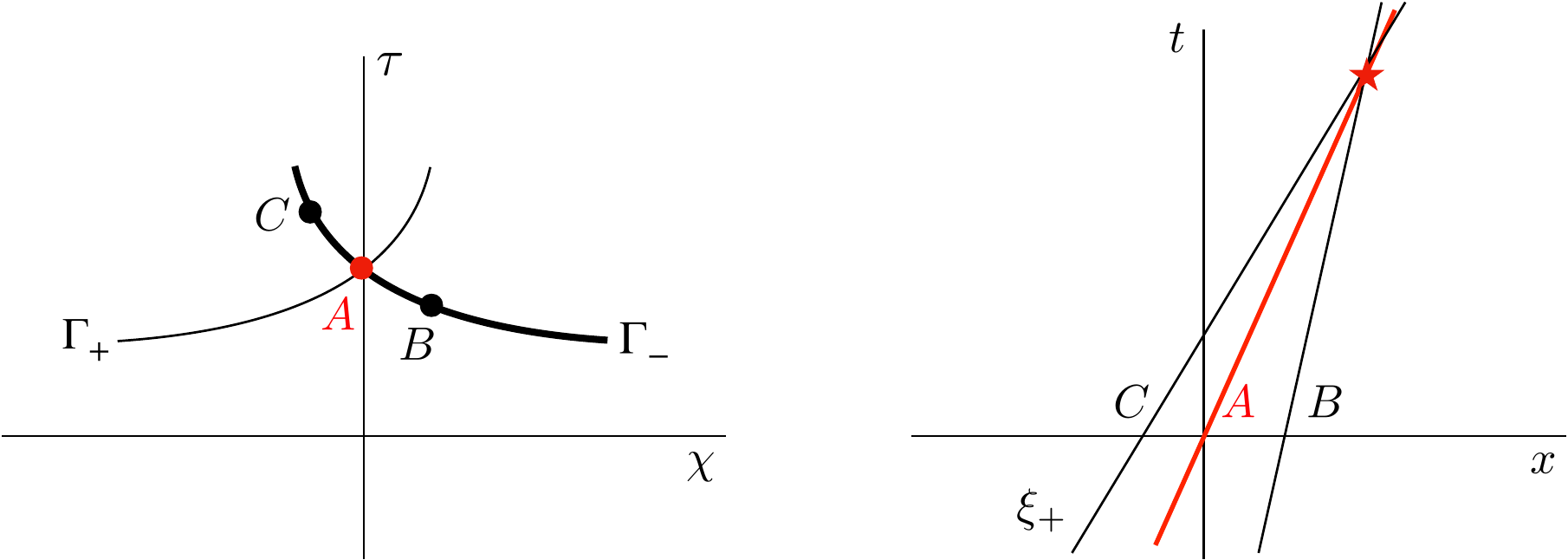}}{\caption{The solution of a simple right moving wave is shown. 
In the left panel the solution is shown in the $(\tau,\chi)$ plane. 
The solution completely lies on a singe characteristic $\Gamma_-$. 
The same solution is shown in the right panel in the physical space. 
The right propagating characteristics $\xi_+$ are straight and non-parallel lines, resulting in their intersection (red star) where a caustic is formed.
}\label{pic1}
\end{figure}

Using the characteristic method sketched above, one can find exact wave solutions of the theory~(\ref{L}).
A simple wave is such a solution of the equation of motion, such that its image in the $(\tau,\chi)$ plane completely sits on one of the characteristics.
An example of a simple wave is shown in Fig.~\ref{pic1}: in the right panel a right moving wave is shown in the physical space $(t,x)$, 
while in the $(\tau,\chi)$ plane the solution is bound to one of the characteristics (in this case $\Gamma_-$), the left panel. 
An image of each characteristic $\xi_+$ in the physical space ($A$, $B$, $C$ in the right panel of Fig.~\ref{pic1}) is one point in the $(\tau,\chi)$ plane.
An important property of such solutions is that the characteristics $\xi_+$ are straight lines in physical space, 
which allows to construct simple wave solutions analytically. 
Note also that the characteristic $A$ propagates with speed $c_s$ as it can be seen from~(\ref{xipm2}). Therefore in the following we will refer to a simple wave for a theory with $c_s<1$ ($c_s>1$) defined in (\ref{cs}) as a subluminal (superluminal) background solution.
Due to the non-linearity of the equation of motion, the characteristics have different inclinations.
This leads to formation of caustics (the red star in the right panel of Fig.~\ref{pic1}), where the second derivatives diverge and the equation of motion is no longer well defined. 
Below we will examine perturbations on top of simple wave solutions and we will show that perturbations become pathological before a caustic is formed.

\section{Perturbations around a background solution}
\label{sec:perturbations}
We will now consider linear perturbations on top of a simple wave, presented in the previous section. 
The second order perturbation of the Lagrangian~(\ref{L}) in 1+1, up to a total derivative, reads,
\begin{equation}
\label{L2}
\mathcal{L}_{2} = \frac12\left(\mathcal{K}_X g^{\mu\nu} + \mathcal{K}_{XX}\nabla^\mu\phi\nabla^\nu\phi\right)\partial_\mu\pi\partial_\nu\pi
-\frac{\gamma}2\left( \phi''\dot\pi^2 -2\dot\phi'\dot\pi\pi' +\ddot\phi\pi'^2 \right),
\end{equation}
where $g^{\mu\nu}=\rm{diag}\{1,-1\}$ is the inverse  Minkowski metric and $\pi$ is the perturbation of the scalar field $\pi \equiv \delta \phi$. 
The Lagrangian~(\ref{L2}) can be rewritten in a compact form,
\begin{equation*}
\label{L2bis}
\mathcal{L}_2 = \frac12 \mathcal{G}^{\mu\nu}\partial_\mu\pi \partial_\nu\pi,
\end{equation*}
where $\mathcal{G}^{\mu\nu}$ is the effective metric for the perturbation $\pi$, given by
\begin{equation}
\label{G}
\mathcal{G}^{\mu\nu} =
\begin{pmatrix}
\mathcal{K}_X+\dot\phi^2 \mathcal{K}_{XX} - \gamma \phi'' &\quad -\dot\phi\phi' \mathcal{K}_{XX} +\gamma \dot\phi'\\
-\dot\phi\phi' \mathcal{K}_{XX} +\gamma \dot\phi' &\quad -\mathcal{K}_X+\phi'^2 \mathcal{K}_{XX} - \gamma \ddot\phi
\end{pmatrix}.
\end{equation}
Note that the first derivatives of the background solution do not change along each characteristic, 
therefore the k-essence part of the expression~(\ref{G}) is constant along any characteristic $\xi_+$ for right-moving simple waves. 
This is true, in particular, for the straight lines $A$, $B$ and $C$ in the right panel of Fig.~\ref{pic1}.
On the contrary, the second derivatives of $\phi$ change with time, and they diverge as the solution approaches a caustic.
We will need therefore to compute the second derivatives of the background solution. 

To do so, let us parametrise the characteristic $\Gamma_-$ in the plane $(\chi,\tau)$, shown by the bold solid curve in the left panel of Fig.~\ref{pic1},
by a parameter $\sigma$ defined along this characteristic: i.e. $\tau=\tau(\sigma)$ and $\chi=\chi(\sigma)$ along this curve. 
Then the value $\Gamma_-(\sigma) = \frac{d\tau(\sigma)}{d\chi(\sigma)}$  as a function of $\sigma$ is also given.
Furthermore, because of the relation~(\ref{charG}), $\sigma$ also parametrises characteristics $\xi_+ = \xi_+(\sigma)$ defined in the $(t,x)$ plane.
Finally, in order to fully describe the solution for a characteristic line in the physical space $x=x(t,\sigma)$, 
we need to specify the position $x$ for each characteristic at some particular moment of time, say $t=0$.
Thus we have a congruence of characteristic right-pointing straight lines parametrised by $\sigma$ as
\begin{equation}
\label{xt}
x(t,\sigma) = \xi_+(\sigma) t + x_0(\sigma).
\end{equation}
Here for each value of the parameter $\sigma$, the characteristic $\xi_+(\sigma) = - \Gamma_-(\sigma) = - \frac{d\tau}{d\chi}$ is given by the slope of $\Gamma_-$ (left panel of Fig.~\ref{pic1}), 
and $x_0(\sigma)$ is the value of $x$ at $t=0$.
The differential of~(\ref{xt}) is easily computed,
\begin{equation*}
	\label{dxt}
	dx(t,\sigma) = \xi_+(\sigma) dt +k(t,\sigma) d\sigma,
\end{equation*}
where $k(t,\sigma) \equiv \left(\frac{dx}{d\sigma}\right)_t$ is the derivative of $x$ with respect to $\sigma$ for a fixed time $t$,
\begin{equation*}
k(t,\sigma) =\frac{d\xi_+}{d\sigma}\, t  + \frac{d x_0}{d\sigma}.
\end{equation*}
Using the above equations it is not difficult to obtain the expressions for the second derivatives of $\phi$,
\begin{equation}
	\label{ddphi}
	\begin{split}
	\phi'' & = \frac{d\chi}{dx} =\left(\frac{d\chi}{d\sigma}\right) k^{-1},\\
	\dot\phi' & = \frac{d\chi}{dt}= \frac{d\tau}{dx} =\left(\frac{d\tau}{d\sigma}\right) k^{-1},\\
	\ddot\phi & = \frac{d\tau}{dt} =- \xi_+\left(\frac{d\tau}{d\sigma}\right) k^{-1}.
	\end{split}
\end{equation}
It immediately follows from Eq.~(\ref{ddphi}) that a caustic forms when  
$k(t,\sigma)=\left(\frac{d\xi_+}{d\sigma}\, t  + \frac{d x_0}{d\sigma}\right)$ vanishes: in this case all the second derivatives diverge, as it can be seen from the above expressions.  
In other words, for each $\sigma$, the time $t_*$ when the caustic forms is determined via the relation
\begin{equation*}
	t_*(\sigma) = -\left(\frac{dx_0}{d\sigma}\right)\left(\frac{d\xi_+}{d\sigma}\right)^{-1}.
\end{equation*}
It should be emphasised  that the shape of the characteristic curve $\Gamma_-$ does not fully determine the values of second derivatives of $\phi$. 
Indeed, $\ddot\phi$, $\dot\phi'$ and $\phi''$  given in~(\ref{ddphi}) also depend on the choice of the function $\chi=\chi(\sigma)$ (or $\tau=\tau(\sigma)$) as well as $x_0=x_0(\sigma)$. This corresponds to the freedom in initial conditions of a wave.
In order to be as general as possible, we do not specify particular form of $\Gamma_-(\sigma)$, $\chi(\sigma)$ or $x_0(\sigma)$.
In what follows we will need, however, relations between the second derivatives. 
Using~(\ref{ddphi}) and (\ref{charG}) one finds,
\begin{equation}
\label{relations}
\dot\phi' = -\xi_+\phi'',\;\;\; \ddot\phi = \xi_+^2 \phi''.
\end{equation}

The characteristics for the effective metric~(\ref{G}) can be defined as long as it is hyperbolic, i.e.
\begin{equation*}
	\left(\mathcal{G}^{01}\right)^2-\mathcal{G}^{00}\mathcal{G}^{11} >0.
\end{equation*}
Cumbersome but a straightforward calculation of the above expression with the use of~(\ref{G}) shows that 
the hyperbolicity of perturbations in the theory~(\ref{L}) on a simple wave solution 
is determined by the hyperbolicity of a pure k-essence (i.e. for $\gamma=0$),
\begin{equation*}
	\left(\mathcal{G}^{01}\right)^2-\mathcal{G}^{00}\mathcal{G}^{11} = \frac{\mathcal{K}^2_X}{c_s^2}.
\end{equation*}
This implies that the condition $c_s^2>0$ should be imposed for the hyperbolicity to hold.
The characteristics for full perturbations $\Xi_\pm$ (not to be confused with the characteristics for pure k-essence $\xi_\pm$ given in~(\ref{xipm2})) can be determined from~(\ref{G}),
\begin{equation*}
	\mathcal{G}^{00}\Xi_\pm^2-2 \mathcal{G}^{01}\Xi_\pm+\mathcal{G}^{11}=0.
\end{equation*}
On top of a simple wave background the solution of the above equation can be written as
\begin{equation}\label{xifull}
	\Xi_{\pm}= \frac{-\tau\chi \mathcal{K}_{XX}\pm \frac{\mathcal{K}_X}{c_s}+\gamma\dot{\phi}' }{\frac{\mathcal{K}_X}{c_s}+\chi^2 \mathcal{K}_{XX}-\gamma \phi''}.
\end{equation}
Recall that $c_s$ in the above expression refers to the speed of propagation in k-essence, defined in~(\ref{cs}).
Also, one can check that setting $\gamma=0$ in the above equation, one recovers the characteristics for k-essence, $\Xi_{\pm} = \xi_\pm$.

The first important property of the solution~(\ref{xifull}) is that the right-propagating  characteristic $\Xi_+$ coincides with those of k-essence, $\xi_+$.
Indeed, using (\ref{relations})---which is valid for the right-moving simple wave---one obtains,
\begin{equation*}
	\Xi_+=\xi_+,
\end{equation*}
i.e. the speed of the signal propagation in the right direction is constant along the straight lines $\xi_+$.

\begin{figure}[t]
\includegraphics[angle=-90,width=0.7\textwidth]{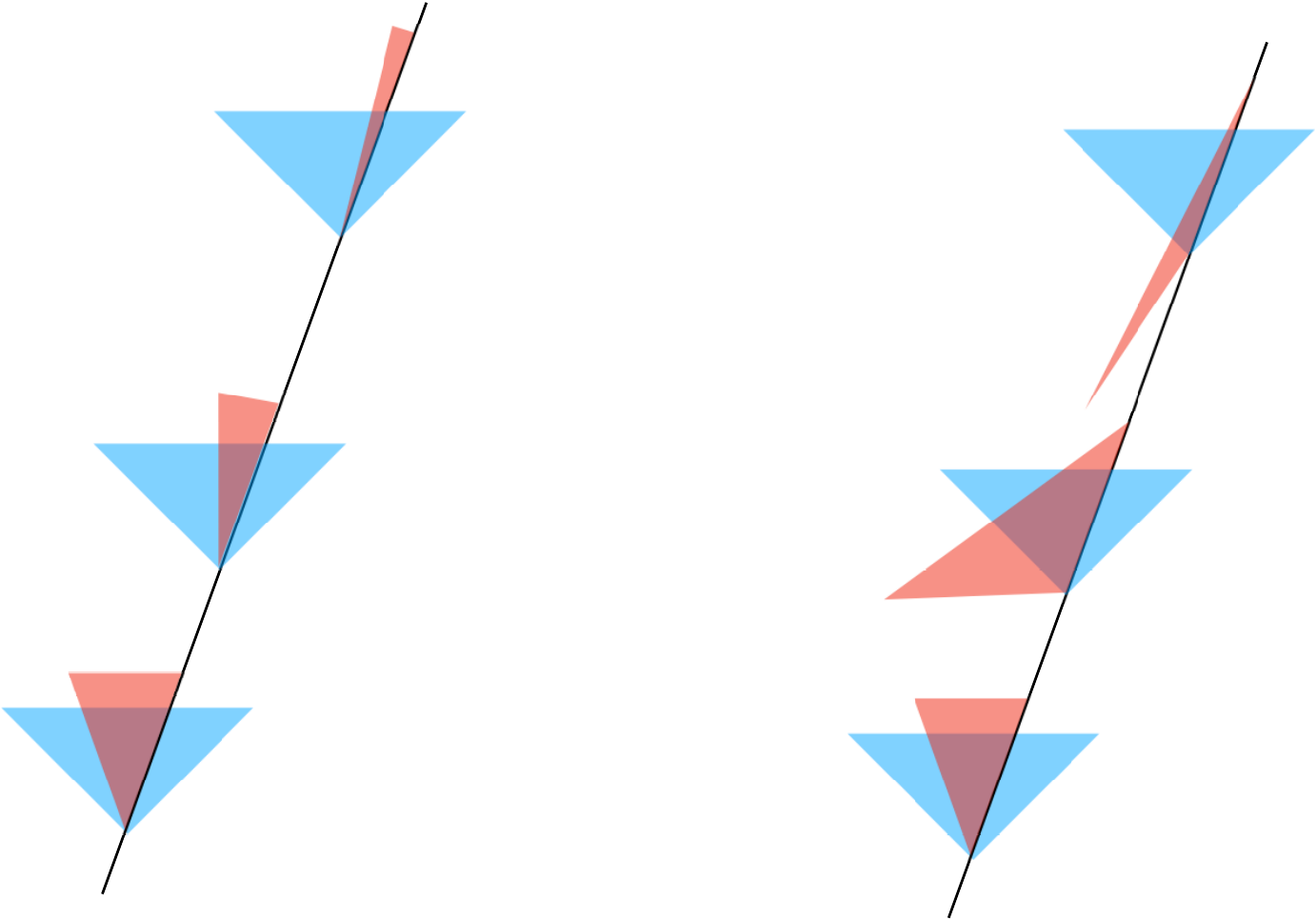}}{\caption{
Future causal cones for normal matter with speed of propagation $c=1$ (blue)
and the cubic galileon (red) for a right-moving subluminal simple wave, prior the caustic formation. 
The evolution of the causal cone is shown along the right propagating characteristic $\Xi_+=\xi_+$.
The left picture corresponds to the case of negative $\gamma \phi''$, while the right picture is for positive $\gamma \phi''$. 
}\label{pic2}
\end{figure}

\begin{figure}[t]
\includegraphics[angle=-90,width=\textwidth]{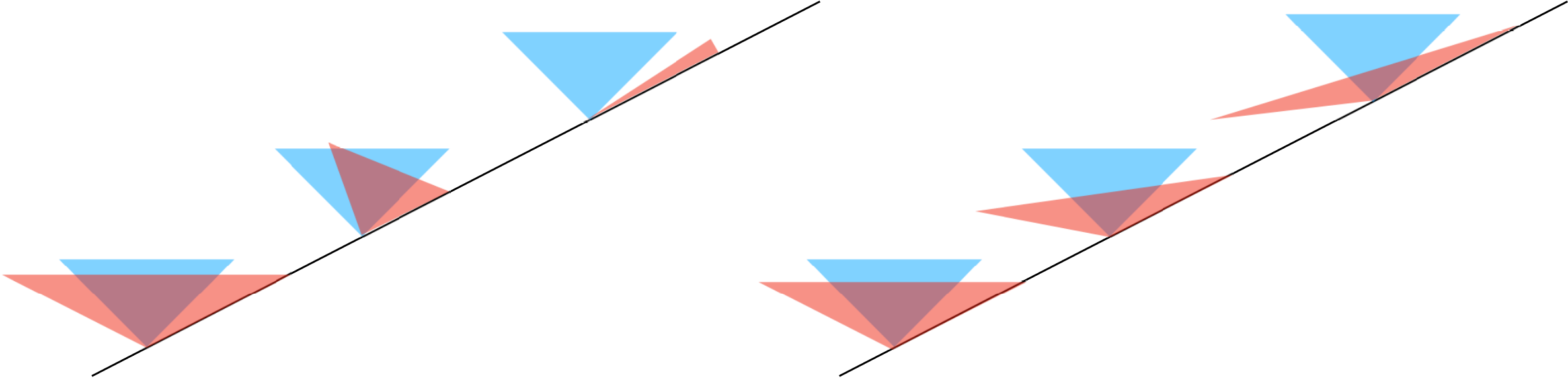}}{\caption{
Future causal cones for normal matter with speed of propagation $c=1$ (blue)
and the cubic galileon (red) for a right-moving superluminal simple wave,  prior the caustic formation. 
The evolution of the causal cone is shown along the right-directed characteristic $\Xi_+=\xi_+$.
The left picture corresponds to the case of negative $\gamma \phi''$, while the right picture is for positive $\gamma \phi''$. 
}\label{pic3}
\end{figure}

On the other hand, the value of $\Xi_-$ changes along the wave propagation, depending on the sign of $\gamma \phi''$.
Close to the caustic, when the second derivatives of $\phi$ diverge, from~(\ref{xifull}) it follows that $\Xi_-\to \xi_+$.
The details of approaching $\Xi_-$ to $\xi_+$ near the caustic, however,  depend on the sign of $\gamma \phi''$.
For negative $\gamma \phi''$, the value of $\Xi_-$ grows starting from negative values $\Xi_- \simeq \xi_-$ far from the caustic (where the terms with second derivatives in (\ref{xifull}) are negligible), crosses zero (so that the characteristic line $\Xi_-$ is pointing towards positive time direction) and becomes positive, finally approaching the value $\xi_+$.
In contrast, for positive $\gamma \phi''$, the characteristic $\Xi_-$ decreases as the solution approaches the caustic, 
crosses $x$ axis (at this moment $\Xi_-$ is infinite), and approaches $\xi_+$, however, pointing in the negative time direction. 
Fig.~\ref{pic2} shows future causal cones for normal matter and the theory~(\ref{L}) on top of a subluminal simple wave.
Fig.~\ref{pic3} is analogous to Fig.~\ref{pic2}, with the difference that it depicts the case of a superluminal simple wave.

In the next section we will demonstrate emergence of ghosts for the perturbations we have studied above. 
Beforehand, let us comment on a possibility to extend our results to include dynamical metric.
For dynamical metric, the expression~(\ref{G}) is modified by an extra term
that originates from the kinetic mixing of the graviton with the scalar field~\cite{Deffayet:2010qz}. 
The extra term is suppressed by the square of the Planck mass and it only affects the ``k-essence'' part of~(\ref{G}). 
Indeed, in order to include the effect of dynamical metric one should add in~(\ref{L2}) the term (to see that one should compare the expression~(\ref{G}) with the one given in~\cite{Deffayet:2010qz}),
\begin{equation}
\label{extra}
-\frac{\gamma^2 X}{2M_{Pl}^2}\left(Xg^{\mu\nu}-2\nabla^\mu\phi\nabla^\nu\phi\right)\partial_\mu\pi\partial_\nu\pi ,
\end{equation}
where $M_{Pl}$ is the Planck mass. 
Note that it is implicitly assumed that the mass scale in~(\ref{L}) is set to unity, therefore the Planck mass in (\ref{extra}) should be given in terms of this scale.
The modification~(\ref{extra}) can be effectively incorporated in~(\ref{L2}) and (\ref{G}) by the change 
$\mathcal{K}_X\to \mathcal{K}_X- \frac{\gamma^2 X^2}{M_{Pl}^2}$ and $\mathcal{K}_{XX}\to \mathcal{K}_{XX}+\frac{2\gamma^2 X}{M_{Pl}^2}$.
Since our study holds for a general non-linear function $\mathcal{K}(X)$, the modification of the perturbation equation due to dynamical metric does not alter the results of this section. 
However, one should also bear  in mind that the evolution of the background solution in Sec.~\ref{sec:background} was also obtained assuming the Minkowski non-dynamical metric.
Indeed, in the case of dynamical metric, a propagating wave backreacts on the metric, causing deformation of the flat background. 
The solution for simple waves we considered in Sec.~\ref{sec:background} is then distorted due to the deformation of the background metric. However, deviations from the flat metric are suppressed by the square of the Planck mass, 
thus can be safely neglected when the characteristic (energy) scales of the solution are smaller than the Planck mass.

\section{Emergence of ghost}
\label{sec:ghosts}

Having found the behaviour of the characteristics for perturbations, one can readily calculate the Hamiltonian density for the perturbations. 
In all the cases depicted in the Figs.~\ref{pic2} and \ref{pic3} the Hamiltonian becomes unbounded from below at some point close to the caustic. 
Indeed, following the standard procedure, we define the conjugate momentum,
\begin{equation*}
p=\frac{\partial\mathcal{L}_2}{\partial \dot\pi} = \mathcal{G}^{00}\dot\pi + \mathcal{G}^{01}\pi'.
\end{equation*}
The Hamiltonian density is then expressed in terms of the conjugate momentum as,
\begin{equation}
	\label{H}
	\mathcal{H}_2 = \frac12\frac{\left(p-\mathcal{G}^{01}\pi'\right)^2}{\mathcal{G}^{00}} - \frac12 \mathcal{G}^{11}\pi'^2.
\end{equation}
One can check that the left panels of Figs.~\ref{pic2} and \ref{pic3} correspond to the change of sign of $\mathcal{G}^{11}$ from negative to positive values as the solution approaches the caustic. 
Change of sign happens at the moment when the characteristic $\Xi_-$ becomes vertical. The appearance of the ``wrong'' sign is related to the time axis being outside the cone. 
For the solutions plotted in the right panels of Figs.~\ref{pic2} and \ref{pic3}, the component $\mathcal{G}^{00}$ changes sign from positive to negative values. 
In this case the sign changes at the moment when the characteristic $\Xi_-$ aligns with $x$ axis, so that after this event the $x$ axis lies inside the cone. This behaviour becomes clear by analysing the cones orientation with respect to the time and space coordinates $t$ and $x$, see details in~\cite{Babichev:2017lmw,Babichev:2018uiw}. Thus, as the solution approaches the caustic singularity, the Hamiltonian becomes unbounded from below, either because $\mathcal{G}^{00}$ or $\mathcal{G}^{11}$ changes sign close to the caustic formation. 
One might conclude that the ghost instability would appears in any case, when the solution is close enough to the caustic formation. 
However, this is not the case, see~\cite{Babichev:2018uiw}. The situation in the right panel of Fig.~\ref{pic2} and the left panel of Fig.~\ref{pic3}
results in the ghost emergence, while in the other cases the unboundedness of the Hamiltonian is merely a result of a poor choice of the coordinate system. 
Indeed, upon the coordinate transformation to some new $\tilde t$ and $\tilde x$, such that the new time axis $\tilde t$ is inside the red cone for the left panel of the Fig.~\ref{pic2}, the Hamiltonian density becomes positive definite all the way down to the caustic singularity. Similarly, by choosing a new $\tilde x$, such that the axis $\tilde x$ is outside both cones in the right panel of Fig.~\ref{pic3}, again, the Hamiltonian density becomes bounded from below. 
Such a trick would not work for the other two cases---the right panel of the Fig.~\ref{pic2} and the left panel of the Fig.~\ref{pic3}---because there is no such a coordinate system that the time axis is inside both cones \emph{and} the $x$ axis is outside of the both cones. 
In this case a physical ghost appears, i.e. vacuum decays into positive energy particles (normal matter) and negative energy particles (galileons), once an interaction between the two species is included.

Let us go through each case in more detail. For subluminal k-essence part of the Lagrangian---i.e. 
when the propagation of perturbations are subluminal in the absence of the cubic galileon---and negative  $\gamma \phi''$ (left panel of Fig.~\ref{pic2}) the future cone for galileon perturbations becomes infinitely narrow as the solution approachs the caustic. One can think of the galileon perturbations in this regime as dust-like. As we already mentioned, in this case no ghost instability appears in spite of the unboundedness of the Hamiltonian density, and also the Cauchy problem is well-posed. 
On the other hand, for negative  $\gamma \phi''$  (the right panel of Fig.~\ref{pic2}), the cone of influence for the galileon perturbations becomes infinitely wide, as the solutions gets close to the caustic singularity. 
First of all, there is a physical ghost in this case, and the solution becomes unstable.
In addition, the Cauchy problem cannot be well-posed simultaneously for the normal matter and the galileon, because there is no Cauchy hypersurface in this case, which is outside of both future cones.
It is worth noting that the appearance of ghost in this case is ``smooth'', i.e. the Hamiltonian density~(\ref{H}) does not become singular at the moment of ghost emergence. 
Indeed, as we discussed above, in this case $\mathcal{G}^{00}$ changes sign when the left characteristic~$\Xi_-$ aligns with the $x$ axis, therefore~(\ref{H}) diverges. However, this particular moment is not related to the appearance of a ghost, but is due to poor choice of coordinates. The ghost appears only some time after $\Xi_-$ crosses the $x$ axis, when the Hamiltonian density does not experience any singular behaviour.

\begin{figure}[t]
\includegraphics[angle=-90,width=0.3\textwidth]{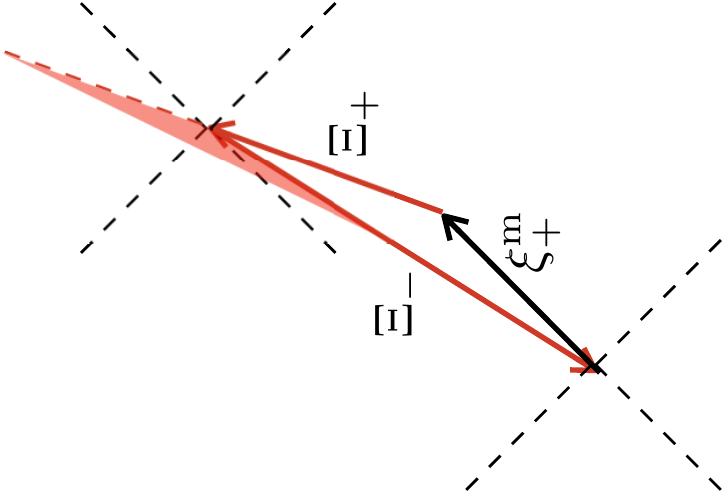}}{\caption{Formation of the closed causal curve is shown for the solution in the right panel of Fig.~\ref{pic2}. Once the future cone for the galileon perturbations opens up enough, so that the characteristic $\Xi_-$ crosses the characteristic $\xi^\text{m}_+$ of the normal matter, closed causal curves appear. If one follows the characteristic $\Xi_-$ back in time, then takes a characteristic of the ordinary matter $\xi^\text{m}_+$ and after that continues along the characteristic $\Xi_+$, a closed causal curve will be formed. 
}\label{pic_ccc}
\end{figure}

For superluminal k-essence part of the Lagrangian, i.e. 
when the propagation of perturbations are superluminal in the absence of the cubic galileon, Fig.~\ref{pic3},
the behaviour of the characteristics is quite different as compared to the subluminal case. 
For negative $\gamma \phi''$ (left panel of the Fig.~\ref{pic3}) the propagation cone for galileon perturbations becomes infinitely thin, similar to the case case shown on the left panel of Fig.~\ref{pic2}, 
with one difference, however, that the galileon cone is now outside of the normal  matter cone,  when the solution approaches to the caustic singularity. So, in contrast to the other case, one cannot interpret the galileon perturbations as dust. 
Moreover, because there is no common time coordinate that would be inside both future cones, the Hamiltonian of the total system is always unbounded from below, as the solution approached the caustic. 
On the other hand, the case of positive $\gamma \phi''$ (the right panel of Fig.~\ref{pic3}) does not lead to the unbounded from below Hamiltonian. In this case the cone of propagation for the galileon becomes infinitely wide, however, upon choosing the hypersurfaces of constant time parallel to $\Xi_+=\xi_+$, the Hamiltonian is positive definite, and the Cauchy problem is well-defined.

We should stress that although the situation shown on the left panel of Fig.~\ref{pic2} and on the right panel of Fig.~\ref{pic3} does not lead to the emergence of a ghost (in spite of the negative Hamiltonian density) it does not mean that the theory is safe from this pathology. Indeed, the situations in the left and the right panels of Figs.~\ref{pic2} and \ref{pic3} are different by the sign of $\gamma \phi''$.
The sign of  $\gamma \phi''$ can be flipped by making the change $\phi \to -\phi$ in the Lagrangian~(\ref{L}), 
such that $\gamma\to -\gamma$. This means that if one of the situations shown in Figs.~\ref{pic2} and \ref{pic3} is realised (say, the left panel),
the other situation (the right panel) also exists, but for different initial conditions.  
Put differently, situations on the left and right panel of each figure differ by the sign of $\dot\phi$, i.e. by the choice of the initial conditions.

It is interesting to point out a connection between the appearance of closed causal curves and the emergence of a ghost\footnote{Closed causal curves in the same model were also studied in~\cite{Evslin:2011vh}. However, the construction of~\cite{Evslin:2011vh} is {\it ad hoc}, since it assumes (without showing it) the existence of a particular non-trivial configuration of the background solution with left and right moving cylinders.}. 
In the situation, depicted in the right panel of Fig.~\ref{pic2}, closed causal curves appear when the future cone of galileon perturbations opens wide, such that the characteristic $\Xi_-$ crosses the characteristic of the ordinary matter $\xi^\text{m}_+$.
In this case a closed causal curve can be formed by taking a triangle, composed of the two characteristic lines $\Xi_-$ and $\Xi_+$, and the characteristic of the ordinary matter $\xi^\text{m}_+$, as it is shown in Fig.~\ref{pic_ccc}. 
However, at the same time, for this configuration the Cauchy problem is not well-posed, and a ghost appears, as it has been discussed above. 
Therefore, from this perspective, the appearance of closed causal curved is plagued with a ghost, therefore time machines can not be created because of the ghost instability. 


\section{Conclusion}
\label{sec:conslusions}
In this paper we investigated  whether an initially healthy solution---with no perturbative instabilities---may dynamically evolve into a configuration containing a ghost. 
In our setup the Lagrangian of the scalar field contains both k-essence and the simplest cubic galileon term, Eq.~(\ref{L}), and the metric is flat and non-dynamical. 
Due to the non-linearity of the k-essence term, a generic non-linear wave solution is prone to the formation of caustics, see Sec.~2 and Ref.~\cite{Babichev:2016hys}. 
Close to a caustic singularity, the second derivatives start to grow and they diverge at the caustic. 
We study perturbations of the scalar field on top of such solutions. 

We showed that before a caustic forms, 
the Hamiltonian for perturbations becomes unbounded from below, independently of the initial conditions for the simple wave.
However, this does not necessarily implies the appearance of an instability~\cite{Babichev:2018uiw}.
In fact, for the solutions shown on the left panel of Fig.~\ref{pic2} and the right panel of Fig.~\ref{pic3} perturbations remain healthy up to the caustic formation. 
In the former case perturbations behave as dust close to the caustic, while in the latter case propagation of perturbations become almost instantaneous, but in both cases perturbations are not ghost-like. 
It is, however, sufficient to change the sign of $\dot\phi$ in the initial condition of background solution for the simple wave, so that one gets the situations shown on the right panel of Fig.~\ref{pic2} and the left panel of Fig.~\ref{pic3} correspondingly, when ghost emerges. 

Thus, generically, even though on the initial hypersurface the perturbations of the galileon are healthy, during dynamical evolution the initially healthy degree of freedom for perturbations becomes a ghost\footnote{A similar study has been considered recently in~\cite{Creminelli:2019kjy}. The difference between our and their approach is that Ref~\cite{Creminelli:2019kjy} considered a high amplitude gravitational wave and they pointed out that perturbations become unstable. 
In our paper, on the contrary, there are no gravitational waves, since the metric is kept flat and non-dynamical all the time, 
and perturbations become ghost-like due to the non-linearity of the galileon Lagrangian itself.}. 
This happens for both subluminal and superluminal simple wave solutions, c.f. Figs.~\ref{pic2} and \ref{pic3}.
However, perturbations are always superluminal when a ghost emerges. 
It is interesting to point out that for a homogeneous cosmological evolution in the same model, 
such an alternation between healthy and ghost-like perturbations  was found to be impossible~\cite{Deffayet:2010qz,Easson:2011zy}, i.e. cosmological evolution either starts or approaches to the boundary between normal and ghost states.
In our case background solutions are not homogeneous, which made it possible for such a conversion of a healthy degree of freedom into a ghost. 
Another important difference between our setup and that of Refs.~\cite{Deffayet:2010qz,Easson:2011zy} is that the gravity in our approach is non-dynamical. However, as we discussed at the end of Sec.~\ref{sec:perturbations}, the modifications introduced by switching on gravity are suppressed by the square of the Planck mass. Thus the limit $M_{Pl}\to \infty$ recovers our setup, while for finite $M_{Pl}$ the corrections are proportional to $M_{Pl}^{-2}$. Keeping in mind that the emergence of a ghost happens before the caustic formation, the modifications due to dynamical gravity can be kept small in the relevant time interval. This argument suggests that taking into account dynamics of gravity does not change our conclusions. 
However, to make this point precise, a separate study is required, which goes beyond our paper.

It should be also stressed that for the solutions analysed in the paper,  perturbations remain hyperbolic all the way down to the formation of the caustic singularity (including the time when ghost exists), i.e. no gradient (or Laplace) instability is present. 
It is also worth to mention that the emergence of a ghost for the configuration shown on the right panel of Fig.~\ref{pic2} is accompanied by the appearance of closed causal curves, as shown in Fig.~\ref{pic_ccc}. 
Said differently, the ghost emerging during evolution forbids the creation of time machines. 

One possible solution to the pathological ghost emergence of the galileon models would be to assume that the theory is effective up to a certain scale, and once second derivatives of the solution become too big, one should not trust such solutions any longer. 
Instead, another, healthy theory should take place for high energies (or, in this case, for large second derivatives of the field), providing a healthy completion of the galileon model.

For future work it would be interesting to made a similar study for more general galileon model and/or to include dynamics of metric perturbations as well. 

\section*{Acknowledgements}
I would like to thank Valery Rubakov, Ignacy Sawicki and Filippo Vernizzi for interesting and stimulating discussions,
and the anonymous referee for valuable comments and questions.
The work was supported by the CNRS/RFBR Cooperation program for 2018-2020 n. 1985 ``Modified gravity and black holes: consistent models and experimental signatures''.


\end{document}